# A constitutive model for brittle granular materials considering the competition between breakage and dilation

Cil, M. B.[1], Hurley, R. C.[2], Graham-Brady, L.[3]

**Abstract**

A constitutive model is presented for brittle granular materials based on a recent reformulation of the breakage mechanics theory. Compared with previous breakage mechanics-based models, the proposed model is improved to capture strain softening towards the critical state following the peak stress observed in dense specimens under shearing, and simultaneous evolution of breakage and dilation. Considering the competition between dilation and particle breakage allows the model to capture breakage-induced reduction in dilatancy and peak strength as confining pressure increases. The influence of the model parameters on the overall material response is described through a detailed calibration procedure based on a benchmark experimental dataset. Comparison of the results of drained triaxial compression experiments on two sands with the predictions of the model indicates that the enriched model successfully captures strain softening in dilatant specimens, the shearing-driven evolution of stress-strain behavior towards the critical state at different confinement levels, the transition from dilatant to compactive volumetric response, and the evolution of particle grading due to distributed breakage events. The proposed framework is capable of qualitatively reproducing the experimentally observed stress-dilatancy-breakage relationship in brittle granular materials in the low pressure regime.

[1] Postdoctoral Researcher, Hopkins Extreme Materials Insitute, Johns Hopkins University, Malone Hall, Suite 140, 3400 North Charles Street, Baltimore, MD 21218
[2] Assistant Professor, Mechanical Engineering, Johns Hopkins University, Malone Hall 117, 3400 North Charles Street, Baltimore, MD 21218
[3] Professor, Civil Engineering, Johns Hopkins University, Latrobe Hall 210, 3400 North Charles Street, Baltimore, MD 21218

# 1. Introduction

The strength and stress-strain behavior of brittle granular materials depend strongly on the initial density and confining pressure (e.g.,Lade and Bopp, 2005, Xiao, *et al.*, 2015, Alshibli and Cil, 2018) as well as the size distribution of particles, which may change irreversibly during loading (Lade, *et al.*, 1996, Wood and Maeda, 2008, Altuhafi and Coop, 2011). At low stresses, inelastic rearrangement of particles primarily dictates the overall response during shearing and results in volumetric dilation or compaction depending on the packing and stress level (Andò, *et al.*, 2013, Alshibli, *et al.*, 2017). At high stresses, particles can begin to crush (Cil and Buscarnera, 2016, Xiao, *et al.*, 2016, Hurley, *et al.*, 2018), resulting in alterations in the stress-strain-volume behavior and physical properties (Lade, *et al.*, 1996, Karatza, *et al.*, 2018, Ciantia, *et al.*, 2019). High-stress conditions leading to particle breakage can be found in many applications, such as in high earth dams, compaction band formation in porous rocks, tunnels, railway ballasts and deep-driven pile foundations (Yasufuku and Hyde, 1995, Das, *et al.*, 2011, Frossard, *et al.*, 2012, Indraratna and Nimbalkar, 2013).

Experimental studies have revealed that particle breakage can cause dense specimens to exhibit reduced dilatancy and even transition to compactive behavior as pressure increases (Bolton, 1986, Xiao, *et al.*, 2016, Yu, 2017a). Another outcome of particle crushing is in changes to the critical (steady) failure state, in which granular materials subjected to shearing continue deforming under a constant shear stress, constant mean effective stress, and constant volume (Schofield and Peter, 1968). Experiments conducted on sands (Bandini and Coop, 2011, Ghafghazi, *et al.*, 2014, Yu, 2017b) and rockfills (Xiao, *et al.*, 2016) indicated that the critical state line in the void ratio and logarithm of mean effective pressure space, *e-log(p)*, descends (i.e., critical state occurs at a lower void ratio) and has a less negative slope when the grading changes due to significant particle crushing. Therefore, incorporating the effects of density, pressure and particle breakage into a constitutive modeling formulation is essential to accurately model critical state phenomena.

Early mathematical formulations in soil mechanics, which generally adopted the principles of the critical state theory (Schofield and Peter, 1968, Gens and Potts, 1988) without directly taking into account particle crushing, provide reasonable predictions in low stress regimes in which breakage has little to no impact on the material response. The predictive capabilities of these models deteriorate at high stresses due to extensive particle breakage. Recent constitutive modeling efforts implicitly incorporated the impact of particle crushing on the mechanical behavior of granular soils using the critical state framework, bounding surface plasticity, and elastoplasticity (Pestana and Whittle, 1999, Cecconi*, et al.*, 2002, Russell and Khalili, 2004, Taiebat and Dafalias, 2008, Yao*, et al.*, 2008). These models are successful in predicting the overall material response; however, they do not directly consider the influence of the initial particle size distribution (PSD) and its evolution in their formulations.

An alternative approach is to explicitly represent particle breakage with indices to establish a link between the evolving gradation and the macroscopic material behavior (Salim and Indraratna, 2004, Einav, 2007a, Kikumoto*, et al.*, 2010, Liu and Zou, 2013, Kan and Taiebat, 2014, Wang and Arson, 2016, Liu*, et al.*, 2017, Zhang and Buscarnera, 2017). In particular, a novel and powerful modeling approach is the theory of breakage mechanics (Einav, 2007a, b), which models the comminution of brittle particulate systems. An essential feature of this modeling approach is the use of micromechanics-inspired internal variables that can be measured using conventional experimental methods (Einav, 2007a, b). Recently, the breakage mechanics theory was extended to predict the critical state of sand under shearing by considering dilation and breakage in different stress regimes (Tengattini, *et al.*, 2016). Although this model can successfully reproduce many key aspects of the behavior of granular materials, including prediction of the breakage- and density-dependent critical state, it is unable to capture the concurrence of breakage and dilation, or the stress softening phenomenon in dilatant specimens.

The main objective of this study is to develop an enhanced constitutive model based on the recent reformulation of breakage mechanics (Tengattini, *et al.*, 2016) for brittle granular materials. Compared with previous breakage-mechanics models, the proposed formulation is improved <u>to capture simultaneous evolution of dilation and breakage and the associated reduction in dilatancy/peak strength with increasing breakage, as well as strain softening observed after peak strength in sheared dense specimens.</u> In Section 2, a detailed description of the model is presented. A step-by-step calibration strategy for the model parameters with reference to an experimental dataset is then illustrated in Section 3. The performance and capabilities of the model is evaluated by comparing model predictions with the results of triaxial compression tests on two sands in terms of stress-strain response, volume change and evolution of PSD in Section 4. Section 5 provides some observations and conclusions.

## 2. An enhanced constitutive model for brittle granular materials

In the following section, the details of the model based on the works of Tengattini, *et al.* (2016), Einav (2007a), Nguyen and Einav (2009) and Rubin and Einav (2011), is presented. The following section then describes two internal state variables (i.e., breakage and porosity), hyper-elastic relations, evolution laws and the yield surface. All analyses are limited to drained behavior and the standard soil mechanics notations and conventions are used by taking compressive stresses and strains as positive, and reporting all stresses as effective stresses.

## 2.1. Notation

In the model formulation, the following standard triaxial test notations in soil mechanics are adopted: mean effective stress $p$, deviatoric (shear) stress $q$, total volumetric strain $\varepsilon_v$ and total deviatoric (shear) strain $\varepsilon_s$, which are computed as:

$$p = \frac{1}{3}\sigma_{ii} \tag{1a}$$

$$q = \sqrt{\frac{3}{2} s_{ij} s_{ij}} \tag{1b}$$

$$\varepsilon_v = \varepsilon_{kk} \tag{1c}$$

$$\varepsilon_s = \sqrt{\frac{2}{3} e_{ij} e_{ij}} \tag{1d}$$

where $\sigma_{ij}$ is the stress tensor, $\varepsilon_{ij}$ is the strain tensor, and the deviatoric stress $s_{ij}$ and strain $e_{ij}$ tensors are given by:

$$s_{ij} = \sigma_{ij} - \frac{\sigma_{kk}}{3}\delta_{ij} \tag{2a}$$

$$e_{ij} = \varepsilon_{ij} - \frac{\varepsilon_{kk}}{3}\delta_{ij} \tag{2b}$$

where $\delta_{ij}$ is the Kronecker delta. The rates of the total volumetric and shear strain increments $(\dot{\varepsilon}_v, \dot{\varepsilon}_s)$ are decomposed into elastic ($\dot{\varepsilon}_v^e, \dot{\varepsilon}_s^e$) and plastic ($\dot{\varepsilon}_v^p, \dot{\varepsilon}_s^p$) terms as:

$$\dot{\varepsilon}_v = \dot{\varepsilon}_v^e + \dot{\varepsilon}_v^p \tag{3a}$$

$$\dot{\varepsilon}_s = \dot{\varepsilon}_s^e + \dot{\varepsilon}_s^p \tag{3b}$$

## 2.2. Internal state variables

### 2.2.1. Breakage index B

In the current formulation, the evolution of the PSD is tracked using the internal variable breakage $B$ (see the definition of $B$ in Figure 1), which was introduced in the continuum breakage mechanics theory (Einav, 2007a) as a practical way of quantifying progressive particle crushing. The breakage variable $B$ measures the relative position of the current PSD $g(x,B)$ between the initial $g_0(x)$ and ultimate $g_u(x)$ PSDs by postulating fractional independence of breakage in different size ranges. $g(x,B)$ can be deduced according to the following relation (Einav, 2007a):

$$g(x,B) = g_0(x)(1-B) + g_u(x)B \tag{4}$$

where $B$ varies from $B=0$ for the initial state without breakage to $B=1$ when the ultimate breakage is reached.

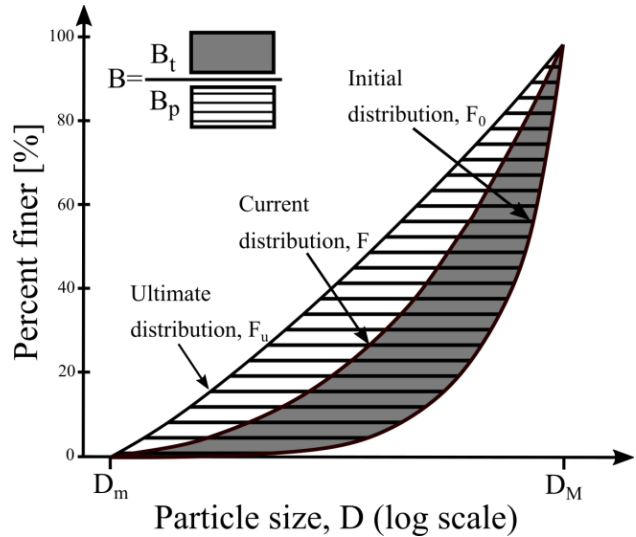

Figure 1. Definition of the internal breakage variable $B$ based on the initial $F_0$, current $F$, and ultimate $F_u$ cumulative particle size distributions. $D_m$ and $D_M$ are the minimum and maximum particle sizes, respectively.

*2.2.2. Porosity*

Porosity $\phi$ is defined as the ratio of the volume of the pore space to the total volume in porous materials. By adopting the ideas of Rubin and Einav (2011) and Tengattini, *et al.* (2016), the porosity $\phi$ is employed as a state variable in the current formulation to model the inelastic deformation of the material. The elastic $\dot{\varepsilon}_v^e$ and plastic $\dot{\varepsilon}_v^p$ volumetric strain rates are then described as (Tengattini, *et al.*, 2016):

$$\dot{\varepsilon}_v^e = -\frac{\dot{V}_s}{V_s} \tag{5a}$$

$$\dot{\varepsilon}_v^p = -\frac{\dot{\phi}}{1-\phi} \tag{5b}$$

where $\dot{\phi}$ is the rate of porosity change, $V_s$ is the solid volume, and $\dot{V}_s$ is the rate of solid volume change. While solid grains are generally assumed to be incompressible in soil mechanics (i.e., the total volume change is related with porosity change), this postulation unrealistically leads to negative porosity values especially in very dense porous materials. Therefore, the above formulation in Equation (5), which was introduced by Tengattini, *et al.* (2016) to eliminate negative porosities, is adopted in this study.

The minimum $\phi_{min}$ and maximum $\phi_{max}$ porosity values depend significantly on the evolution of PSD associated with particle breakage since a wider particle gradation with a sufficient number of smaller fragments facilitates denser packing by permitting smaller fragments to fill available pore space. To capture the breakage dependency of two porosity limits ($\phi_{min}, \phi_{max}$), the following relations, which are modified versions of the expressions proposed by Rubin and Einav (2011), are used:

$$\phi_{min} = \phi_l (1-B)^l \exp(-Bl) \tag{6a}$$

$$\phi_{max} = \phi_u (1-B)^u \exp(-Bu) \tag{6b}$$

where $\phi_l$ and $\phi_u$ are the minimum and maximum porosity limits without any breakage, respectively, and $u$ and $l$ are the coefficients that control the evolution of these limit porosities. The exponential terms in Equation (6) are added to the expressions of Rubin and Einav (2011) to mitigate the overprediction of volumetric compaction at high breakage values based on the observations made during the calibration/validation of the model. The comparison of these breakage-dependent porosity limit relations against the experimental results of a sand mixture reported by Youd (1973) is displayed in Figure 2. Additionally, the relative porosity index $\tau$, which is also grading-dependent, is employed in the proposed formulation and is given by (Rubin and Einav, 2011):

$$\tau = \frac{\phi_{\max} - \phi}{\phi_{\max} - \phi_{\min}} \tag{7}$$

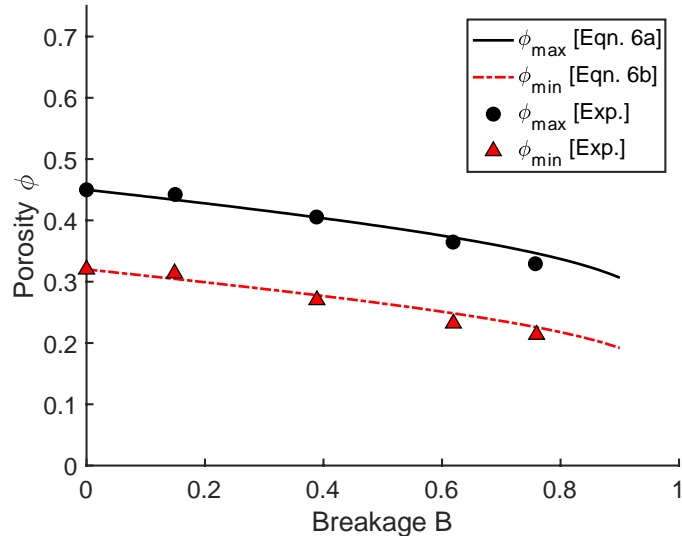

Figure 2. Variation of the minimum $\phi_{\min}$ and maximum $\phi_{\max}$ porosity values as a function of breakage $B$ predicted by Equations (6a) and (6b) using parameters $\phi_l = 0.32$, $\phi_u = 0.45$, $u = 0.12$, and $l = 0.16$. The experimental data is presented by Youd (1973) and the breakage values are digitized from the results reported by Tengattini, et al. (2016) who computed $B$ values based on the reported PSDs by Youd (1973).

*2.3. Hyper-elastic formulation*

The following pressure-dependent Helmholtz free energy potential (Nguyen and Einav (2009)) is adopted to capture the non-linear elastic response of brittle granular materials:

$$\Psi = (1 - \vartheta B)\left( p_r \frac{2A^3}{3\bar{K}} + p_r \frac{3}{2}\bar{G}A\left(\varepsilon_s^e\right)^2 \right) \tag{8}$$

where

$$A = \left(\frac{1}{2}\bar{K}\varepsilon_v^e + 1\right) \tag{9}$$

where $p_r$ is a reference pressure (here taken as 1 kPa) and $\bar{K}$ and $\bar{G}$ are the non-dimensional material constants. $\vartheta$ is the grading index introduced by Einav (2007a) as a result of a statistical homogenization procedure to incorporate the influence of the evolving PSD. $\vartheta$ represents the relative distance between the initial PSDs based on the second order moments of these gradings and is computed as (Einav, 2007a):

$$\vartheta = 1 - \frac{\int_{D_m}^{D_M} g_u(x) x^2 dx}{\int_{D_m}^{D_M} g_0(x) x^2 dx} \tag{10}$$

Based on the energy potential given in Equation (8), the following constitutive relations are derived for the triaxial stresses $(p, q)$ and the breakage energy $E_B$:

$$p = \frac{\partial \Psi}{\partial \varepsilon_v^e} = (1 - \vartheta B)\left(3 p_r A^2 + p_r \frac{3}{4}\bar{G}\bar{K}\left(\varepsilon_s^e\right)^2\right) \tag{11a}$$

$$q = \frac{\partial \Psi}{\partial \varepsilon_s^e} = (1 - \vartheta B)\left(3 p_r \bar{G} \varepsilon_s^e\right) \tag{11b}$$

$$E_B = -\frac{\partial \Psi}{\partial B} = \vartheta\left(p_r \frac{2A^3}{3\bar{K}} + p_r \frac{3}{2}\bar{G}A\left(\varepsilon_s^e\right)^2\right) \tag{11c}$$

The breakage energy $E_B$ physically represents the energy required for particle crushing to shift the PSD from the initial to ultimate state (Einav, 2007a).

*2.4. Yield surface and flow rules*

In constitutive model formulation, one of two approaches are typically employed to obey the second law of thermodynamics in isothermal conditions (i.e., the rate of material dissipation, $\Phi \geq 0$). The first approach, which is adopted by Tengattini, *et al.* (2016) and discussed in detail by Houlsby and Puzrin (2007), is to devise the yield surface and flow rules from a dissipation potential. The second approach, which is adopted in this study and by Rubin and Einav (2011), involves directly proposing the yield function and flow rules, and imposing restrictions to ensure that the total dissipation rate $\Phi$ is non-negative.

The following yield function, $y$, in mixed stress-energy space, which is a modified version of the function derived by Tengattini, *et al.* (2016), is proposed:

$$y = \left(\sqrt{\frac{E_B}{E_C}}(1-B) - \gamma\tau\right)\sqrt{\frac{E_B}{E_C}(1-B)^2} + \left(\frac{q}{(M_d + M)p}\right)^2 - 1 \leq 0 \qquad (12)$$

where $E_C$ is the critical breakage energy that controls the onset of particle breakage under isotropic loading conditions, $M$ is the ratio of the deviatoric stress $q$ to mean effective stress $p$ at the critical state, the parameter $\gamma$ regulates the dilatant behavior and should vary between 0 and 1 to ensure that the total dissipation rate is still positive during dilation. Compared to the yield function devised by Tengattini, *et al.* (2016), the yield surface given in Equation (12) is modified by multiplying the first term with $\sqrt{E_B/E_C}(1-B)^2$ to eliminate non-convexity at low confining pressures. $M_d$ is also introduced in Equation (12) to capture stress- and density-dependent peak strength, and strain softening observed in dilatant specimens. $M_d$ represents the additional stress ratio due to dilatancy relative to the critical state stress ratio $(M = q_{cs}/p_{cs})$ and is computed as follows:

$$M_d = \gamma \langle \tau - \tau_{cs} \rangle^2 \left(\frac{6\sin(\theta_p)}{3-\sin(\theta_p)} - M\right), \qquad (13)$$

where $\theta_p$ represents the peak state friction angle (i.e., here assumed as the summation of the dilatancy angle $\theta_d$ and the critical state friction angle $\theta_{cs}$), and $\tau_{cs}$ is the critical state relative porosity during dilation. The terms in Eqn. (13) are evaluated as:

$$\theta_p = \theta_d + \theta_{cs}, \quad \theta_d = \frac{\pi}{15}, \quad \theta_{cs} = \sin^{-1}\left(\frac{3M}{6+M}\right), \tag{14a}$$

$$\tau_{cs} = \sqrt{\frac{E_B}{E_C}\frac{(1-B)}{\gamma}}. \tag{14b}$$

The value of $\theta_d$, which represents the maximum dilatancy angle, is taken as 12° (i.e., $\pi/15$ rad), based on previous experimental data obtained on various sands (Bolton, 1986, Alshibli and Cil, 2018). The use of Macaulay brackets $\langle\ \rangle$ (i.e., $\langle x \rangle = (x + |x|)/2$) implies that $M_d$ is only defined for dilatant behavior (i.e., if $\tau - \tau_{cs} > 0$, $\langle \tau - \tau_{cs} \rangle = \tau - \tau_{cs}$, otherwise $\langle \tau - \tau_{cs} \rangle = 0$, and $M_d = 0$) and gradually approaches zero to capture shearing-driven evolution towards the critical state. The value of $M_d$ decreases as porosity $\phi$ or pressure level increases, which is in agreement with previous experimental findings on sands (Bolton, 1986, Alshibli and Cil, 2018).

The following evolution laws for breakage $\dot{B}$, porosity $\dot{\phi}$, and plastic shear strain $\dot{\varepsilon}_s^p$, inspired from the work of Tengattini, *et al.* (2016), are used in the model:

$$\dot{B} = \begin{cases} 2\lambda \left\langle \sqrt{\frac{E_B}{E_C}}(1-B) - \gamma_B \tau \right\rangle \frac{(1-B)\kappa\tau}{\sqrt{E_B E_C}} & \text{if } |F| > 0 \\ 0 & \text{if } |F| = 0 \end{cases} \tag{15a}$$

$$\dot{\phi} = 2\lambda \left(\sqrt{\frac{E_B}{E_C}}(1-B) - \gamma\tau\right) \sqrt{\frac{E_B}{E_C}} \frac{(1-B)(1-\tau H(F))}{E_\phi} \tag{15b}$$

$$\dot{\varepsilon}_s^p = 2\lambda \frac{q}{\left((M_d + M)p\right)^2} \tag{15c}$$

where

$$F = \sqrt{\frac{E_B}{E_C}}(1-B) - \gamma\tau \tag{16}$$

where $\lambda$ is the non-negative multiplier that can be computed using the consistency condition to the yield domain as in classical elasto-plasticity, $\gamma_B$ regulates the initiation of particle breakage (i.e., $0 \leq \gamma_B \leq \gamma$), the parameter $\kappa$ controls the crushability of the material, and $E_\phi$ is the stress conjugate to $\dot{\phi}$ ($E_\phi = -p/(1-\phi)$, (Tengattini, et al., 2016)). Breakage growth $\dot{B}$ is assumed to eventually cease once the critical state is reached (i.e., $\dot{B} = 0$ when $F = 0$). $H(F)$ is the Heaviside (unit) step function whose value depends on the sign of the $F$ function (i.e., $H(F) = 0$ if $F < 0$ or $H(F) = 1$ if $F \geq 0$).

In order to allow simultaneous evolution of breakage and dilation, the flow rules given in Equation (15a-b) do not include the coupling angle adopted by Tengattini, et al. (2016) (i.e., Equation (15a) does not involve $H(F)$). Instead, the parameter $\kappa$ is introduced in Equation (15a) to regulate the evolution of breakage. $\tau$ and $(1 - \tau H(F))$ are used in Equations (15a) and (15b) to maintain porosity between the two breakage-dependent porosity limits ($\phi_{min}$ and $\phi_{max}$). $M_d$ is added to Equation (15c) since the stress ratio goes above the critical state line during dilation. These modifications are proposed to qualitatively reproduce the experimentally-observed stress-dilatancy-breakage relationship at low stresses.

It is possible to prove the non-negativity of the total dissipation rate $\Phi$, which is given by (Tengattini, et al., 2016):

$$\Phi = E_B \dot{B} + E_\phi \dot{\phi} + q \dot{\varepsilon}_s^p \geq 0 \qquad (17)$$

In the case of compaction, consulting Equation (15) (i.e., $E_B \dot{B} \geq 0$, since $F \geq 0$, $E_\phi \dot{\phi} \geq 0$, and $q \dot{\varepsilon}_s^p \geq 0$), the sum of all terms in Equation (17) is found to be non-negative. In the case of dilation, the dissipation rate $\Phi$ can be expressed by combining Equations (12), (15), and (17):

$$\Phi = E_B \dot{B} + 2\lambda \left[ 1 + F \sqrt{\frac{E_B}{E_C}} (1-B) B \right] \geq 0 \tag{18}$$

The sum of the terms in Equation (18) is non-negative (i.e., $E_B \dot{B} \geq 0$, $-1 \leq F \leq 0$, $0 \leq B \leq 1$, and $0 \leq \sqrt{E_B / E_C} (1-B) \leq 1$, so $-1 \leq F \sqrt{E_B / E_C} (1-B) B \leq 0$). Consequently, the non-negativity of the total dissipation rate $\Phi$ in the proposed model is always guaranteed, thus satisfying the second law of thermodynamics.

In order to demonstrate the impact of the proposed modifications, the drained triaxial compression response and yield surface evolution for a dense specimen with an initial relative porosity of $\tau_0 = 1$ under confining pressures $\sigma_c = 0.5$ MPa, 1.8 MPa, and 7 MPa are illustrated in Figure 3. The specimen subjected to shearing at $\sigma_c = 0.5$ MPa exhibits a strain softening response (i.e., the shrinkage of the yield domain (Figure 3a) with dilation (Figure 3e)) after reaching a peak strength (Figure 3d). In the second case shown in Figure 3b, the specimen is sheared at a higher confinement ($\sigma_c = 1.8$ MPa), and the occurrence of particle breakage (Figure 3f) results in reduction in dilation (Figure 3e), and peak strength (Figure 3d). Shearing the specimen at $\sigma_c = 7$ MPa resulted in compaction (Figure 3e) and considerable breakage (Figure 3f) (i.e., expansion of the yield surface (Figure 3c)). As summarized in the results shown in Figure 3, in addition to well-established features of breakage-mechanics based models, the improvements introduced in this study allow the new model to capture the peak stress followed by strain softening towards the critical state in dense specimens (Figure 3d), concurrent evolution of breakage and dilation (Figure 3e,f), and reduction in dilatancy and peak strength with increasing breakage (Figure 3d,e).

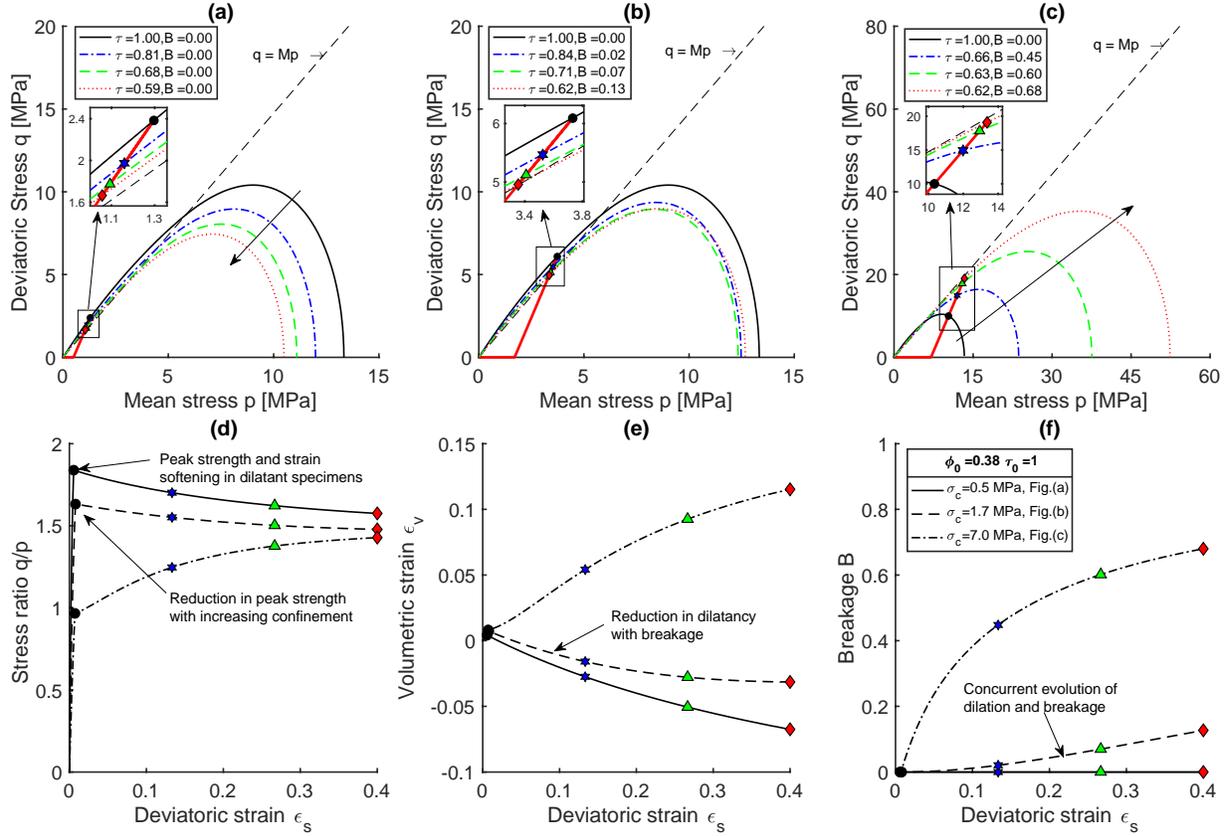

Figure 3. Drained triaxial compression response and evolution of the yield surface for a dense specimen with an initial relative porosity of $\tau_0 = 1$ subjected to drained triaxial compression at confining pressures (a) $\sigma_c = 0.5$ MPa; (b) $\sigma_c = 1.8$ MPa; (c) $\sigma_c = 7$ MPa; (d) stress ratio-deviatoric strain response; (e) volumetric strain behavior; (f) breakage B growth. Each symbol shown in results in (d-f) corresponds to a particular stress state in yield surfaces shown in (a-c).

## 3. Model calibration and validation

The calibration of model parameters and performance analysis of the predictive capabilities of the model are carried out using the results of drained triaxial compression tests performed on Kurnell sand (Russell, 2004, Russell and Khalili, 2004) and Cambria sand (Yamamuro, 1993, Yamamuro and Lade, 1996) under a wide range of confining stresses. In the following section, a detailed calibration strategy for the primary model parameters are described and discussed for Kurnell sand.

*3.1. Calibration of model parameters*

*3.1.1. Physical index properties*

The initial $g_0(x)$ and ultimate $g_u(x)$ PSDs (i.e., probability density functions) can be expressed using the following equations (Einav, 2007a):

$$g_0(x) = \frac{(3-\beta)x^{(2-\beta)}}{D_M^{3-\beta} - D_m^{3-\beta}} \tag{19a}$$

$$g_u(x) = \frac{(3-\alpha)x^{(2-\alpha)}}{D_M^{3-\alpha} - D_m^{3-\alpha}} \tag{19b}$$

where $D_m$ and $D_M$ are the minimum and maximum particle sizes of the corresponding distribution, $\beta$ and $\alpha$ are the coefficients calibrated to capture the initial and ultimate PSDs, respectively. The initial PSD data of Kurnell sand measured through sieve analysis in the study of Russell (2004) is captured using Equation (19a). The ultimate PSD is assumed as a fractal grading (Sammis, *et al.*, 1987, Marone and Scholz, 1989, Ben-Nun and Einav, 2010). The grading index $\vartheta$ is then computed using Equation (10). The same analysis is repeated for Cambria sand based on the PSD data reported in Yamamuro (1993). The initial and predicted ultimate PSDs of two sands are shown in Figure 4 and all the parameters are listed in Table 1. The effect of the fractal coefficient on the ultimate PSD is illustrated in Figure 4b, and a fractal coefficient $\alpha$ of 2.6 is assumed in further analysis for both sands based on previous observations (Sammis, *et al.*, 1987, Marone and Scholz, 1989). The minimum $\phi_{min}$ and maximum $\phi_{max}$ porosity limits for unbroken Kurnell sand are reported as $\phi_l = 0.375$ and $\phi_u = 0.479$ by Russell (2004). The coefficients $l$ and $u$ calibrated for the experimental data (i.e., the sand mixture data reported by Youd (1973)) shown in Figure 2 are used for both sands.

Table 1. Parameters used to capture the initial and ultimate PSD of Kurnell and Cambria sands

| Material | Initial PSD | | | Ultimate PSD | | | Grading index |
|---|---|---|---|---|---|---|---|
| | $D_M$ [mm] | $D_m$ [mm] | $\beta$ | $D_M$ [mm] | $D_m$ [mm] | $\alpha$ | $\vartheta$ |
| Kurnell sand | 0.47 | 0.15 | 2.0 | 0.47 | 0.002 | 2.6 | 0.604 |
| Cambria sand | 2.0 | 0.9 | -1.5 | 2.0 | 0.002 | 2.6 | 0.749 |

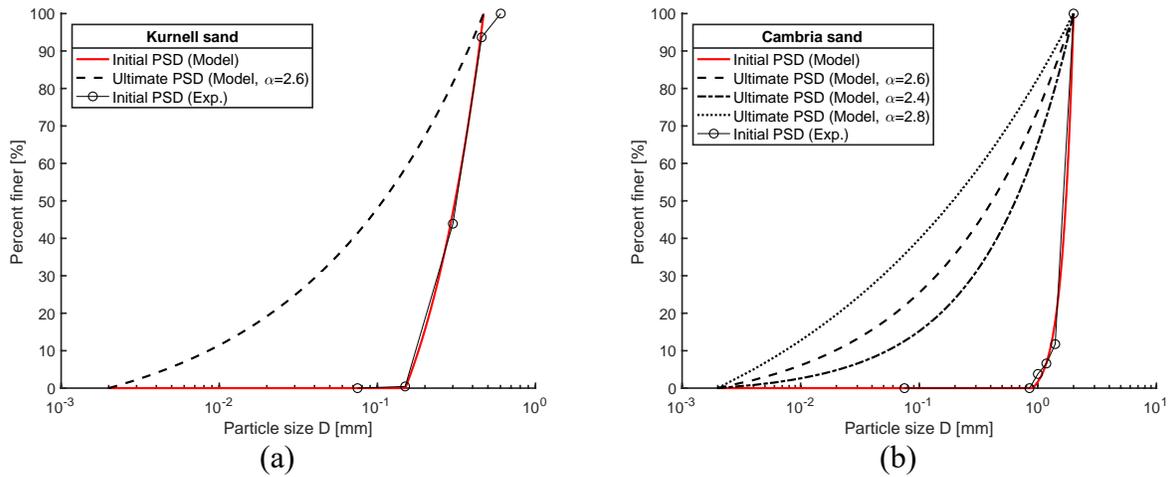

(a)  (b)

Figure 4. Initial and predicted ultimate PSDs of (a) Kurnell sand (Russell, 2004) and (b) Cambria sand (Yamamuro, 1993). The initial PSD curves are fitted by Equation (19a) and a fractal ultimate PSD is postulated for both sands. (b) Effect of the fractal coefficient $\alpha$ on the ultimate PSD is illustrated.

*3.1.2. Mechanical model parameters*

The frictional strength parameter $M$, which corresponds to the stress ratio between the deviatoric stress $q_{cs}$ and the mean effective stress $p_{cs}$ at the critical state, is computed using the critical state friction angle of $\theta_{cs} = 36.3°$ reported by Russell and Khalili (2004) as follows:

$$M = \frac{6\sin\theta_{cs}}{3-\sin\theta_{cs}} \tag{20}$$

The dimensionless elastic constants $\bar{K}$ and $\bar{G}$ can be determined using isotropic and triaxial compression experiments. The critical breakage energy $E_C$ can be calibrated through an oedometric or isotropic compression test. $E_C$ is determined for Kurnell sand using the oedometric compression test data reported by Russell (2004).

The volume change result of a drained triaxial compression experiment is required to determine the parameter $\gamma$ that governs dilatancy. The predicted stress-strain and volume deformation behaviors for different $\gamma$ values are displayed in Figure 5 along with the result of a drained triaxial compression test on Kurnell sand at a confining pressure $\sigma_c$ of 760 kPa. Figure 5 shows that the material exhibits a more dilative response and a slight rise in the peak deviatoric stress as $\gamma$ increases. The value of $\gamma$ is selected as 0.93 to adequately capture the volumetric strain (i.e., total dilation) at the critical state. It is important to note that the parameter $\gamma$ is calibrated through a single experiment, which is eventually employed to predict the critical state over a range of densities and pressures. This approach provides significant advantages compared to the models that require the complete description of the critical state as input for model formulation.

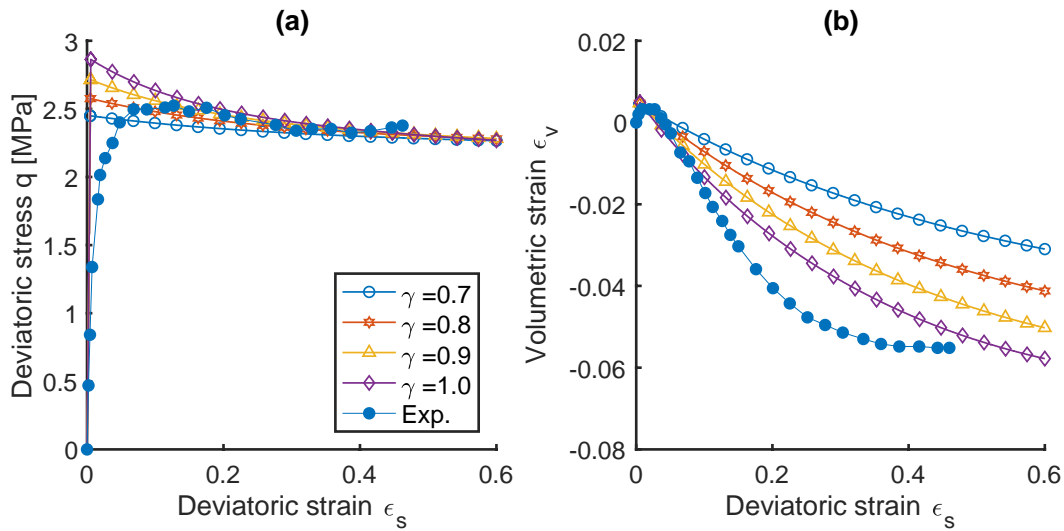

Figure 5. Comparison of the result of a drained triaxial compression experiment with model predictions for different values of $\gamma$ (initial porosity $\phi_0 = 0.40$ and confining pressure $\sigma_c = 760$ kPa). (a) deviatoric stress-strain relationship, and (b) volume change response.

The parameter $\gamma_B$ in Equation (15a) controls the critical energy threshold for the onset of particle breakage under shearing. The influence of the parameter $\gamma_B$ on stress-strain response, volume change, and PSD is presented in Figure 6. The model predictions for a range of $\gamma_B$ values are compared with the results of a triaxial test conducted on Kurnell sand under a confining pressure $\sigma_c$ of 1417 kPa. Reduction in $\gamma_B$ dampens the amount of dilation and results in more particle crushing as seen in predicted PSD changes. The parameter $\gamma_B$ is chosen as 0.68 to reasonably capture both the volumetric response and evolved PSD. The model cannot capture the stress-strain response since the final deviatoric stress in the experiment is relatively lower than the critical stress state estimated based on the average critical state angle $\theta_{cs}$ of all experiments (Figure 6a). The model also predicts a small peak in the deviatoric stress-strain behavior due to dilation (Figure 6a,b).

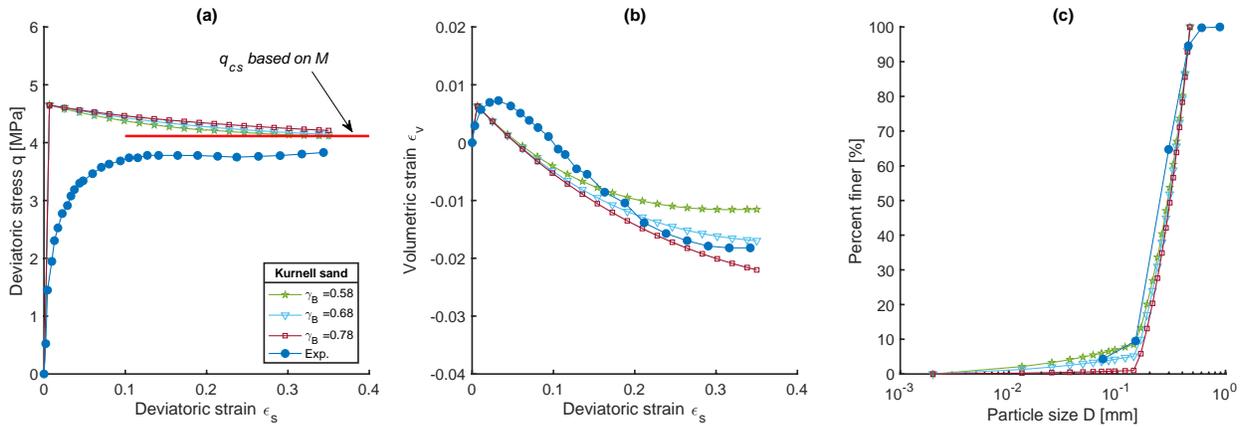

Figure 6. Comparison of the results of a drained triaxial compression experiment with model predictions for different values of $\gamma_B$ (initial porosity $\phi_0 = 0.40$ and confining pressure $\sigma_c = 1417$ kPa ). (a) deviatoric stress-strain relationship, (b) volumetric strain response, and (c) evolution of PSD.

The parameter $\kappa$ given in Equation (15a) regulates the evolution of breakage. The triaxial test result of Kurnell sand at a confining pressure $\sigma_c$ of 7800 kPa is used to illustrate the calibration process for $\kappa$ (Figure 7). As shown in Figure 7, while the parameter $\kappa$ primarily controls breakage growth, it also impacts the stress-strain and volume change behaviors. Increase in the value of $\kappa$ causes less compactive volume change (Figure 7b) and slower evolution of the deviatoric stress and volumetric strain towards the critical state (Figure 7a). Since $\kappa$ influences multiple aspects of the material response, its value is thus chosen as $\kappa = 0.05$ for Kurnell sand to reasonably capture the stress-strain behavior, volume change, and PSD evolution. All the calibrated model parameters for Kurnell and Cambria sands are listed in Table 2.

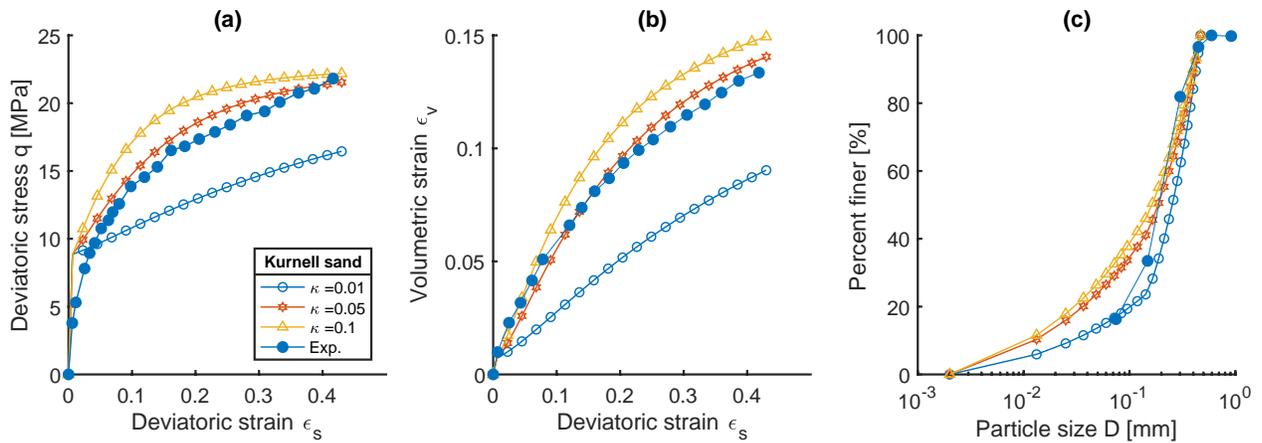

Figure 7. Comparison of the results of a drained triaxial compression experiment with model predictions for different values of the parameter $\kappa$ (initial porosity $\phi_0 = 0.40$ and confining pressure $\sigma_c = 7800$ kPa). (a) deviatoric stress-strain relationship, (b) volumetric strain response, and (c) evolution of PSD.

Table 2. Mechanical and index model parameters for Kurnell and Cambria sands

| Model parameters | Kurnell sand | Cambria sand | Material index properties | Kurnell sand | Cambria sand |
|---|---|---|---|---|---|
| $\bar{K}$ | 3600 | 4000 | $\phi_l$ | 0.375 | 0.334 |
| $\bar{G}$ | 4750 | 5500 | $\phi_u$ | 0.479 | 0.442 |
| $E_C$ [Pa] | 70e3 | 100e3 | $l$ | 0.16 | 0.16 |
| $\kappa$ | 0.05 | 0.06 | $u$ | 0.12 | 0.12 |
| $M$ | 1.475 | 1.353 | $\vartheta$ | 0.604 | 0.749 |
| $\gamma$ | 0.93 | 0.80 | | | |
| $\gamma_B$ | 0.68 | 0.58 | | | |

## 4. Validation of the model

After calibrating model parameters using reference test data, the predictive capabilities of the constitutive model are examined by comparing model predictions with a set of benchmark laboratory experiments conducted over a wide range of stress levels. The model simulations and results of triaxial compression tests on Kurnell sand reported by Russell (2004) are presented in Figure 8. For each model simulation, the initial porosity $\phi_0$, initial PSD $g_0(x)$ and confining pressure $\sigma_c$ are provided as input information to predict the stress-strain-volume change response and evolution of PSD of specimens under confining pressures in the range of 760 kPa to 7800 kPa. The experiments and model simulations are denoted by continuous lines with solid and open symbols, respectively.

With only a set of parameters, the model captures several key characteristics of the behavior of Kurnell sand. The model's use of the parameter $M_d$ allows it to predict peak strength in dilatant specimens and its reduction as confining pressure increases (Figure 8a). The comparison of the stress-strain results of experiments with model predictions are shown in Figures 8b and 8d. The triaxial experimental conducted at confining pressures of 760 kPa, 1417 kPa and 7800 kPa are employed for calibration (Figure 8b). The proposed model successfully captures the stress-strain (Figure 8b) and volume change (i.e., transitioning from dilation to compaction (Figure 8d)) responses of specimens subjected to different confining pressures. The predicted breakage $B$ values increase gradually as confinement increases (Figure 8c). The comparison of the PSDs predicted by the model (Figure 8e) with that measured in experiments (Figure 8f) indicates that the model slightly overestimates the amount of particle fragmentation at high pressures. With the proposed changes, the enhanced model is capable of capturing the concurrent evolution of breakage (Figure 8c,e,f) and dilation (Figure 8d) (see the response of the specimens sheared under confining pressures $\sigma_c$ of 1010 kPa and 1417 kPa).

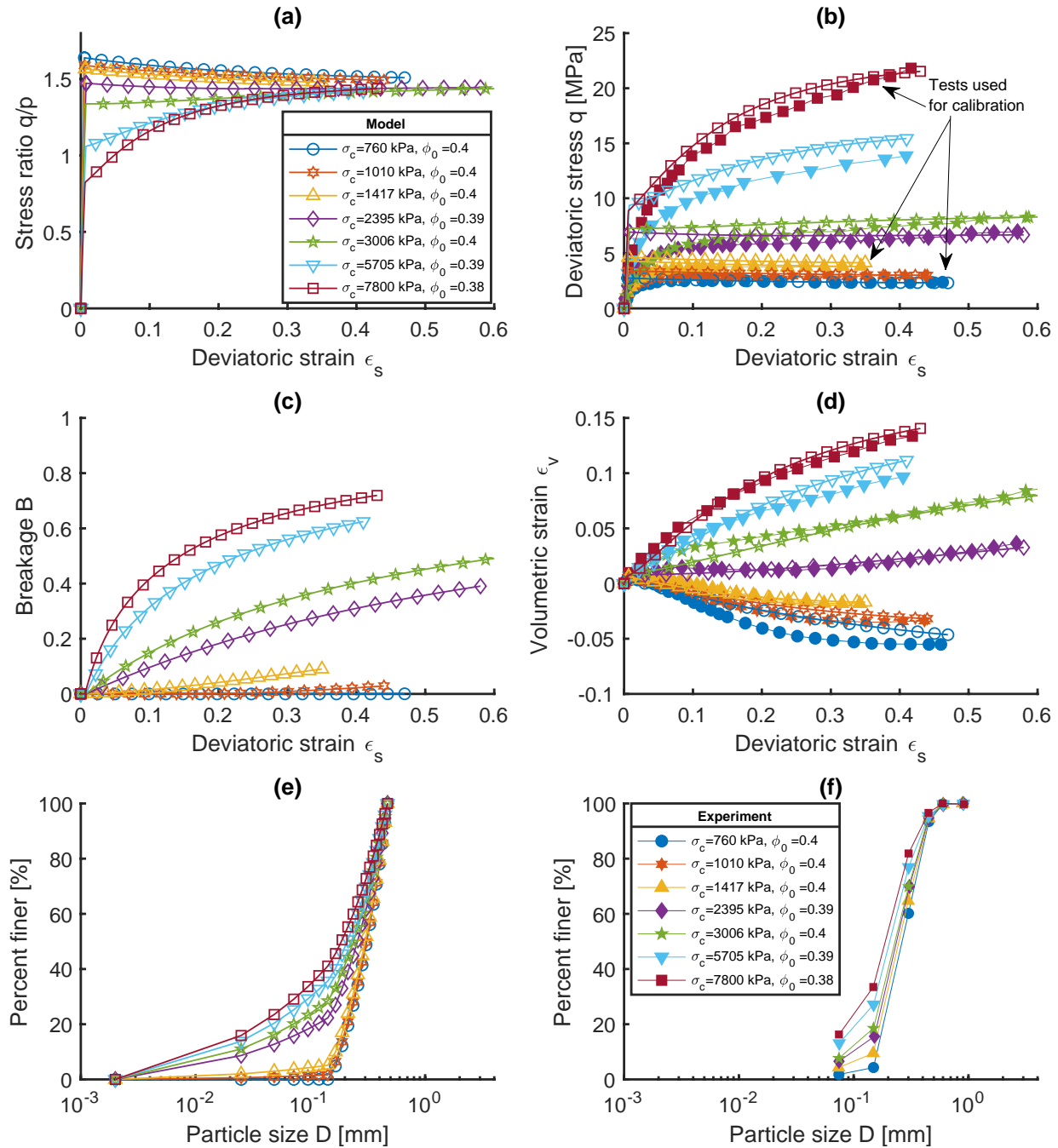

Figure 8. Drained triaxial compression test results (Russell, 2004) and model predictions for Kurnell sand under confining pressures between 760 kPa and 7800 kPa. (a) model prediction of stress ratio versus deviatoric strain response; (b) comparison of the deviatoric stress-strain relationship of experiments (curves with solid symbols) with model predictions (curves with open symbols); (c) model prediction of breakage $B$ growth; (d) comparison of the volume change

response of experiments (curves with solid symbols) with model predictions (curves with open symbols); (e) model prediction of cumulative PSD; (f) cumulative PSD data measured after experiments.

The second experimental dataset employed in model validation involves the results of drained triaxial compression tests on Cambria sand under confining pressures ranging from 2.11 MPa to 15 MPa, reported by Yamamuro (1993) and Yamamuro and Lade (1996). All the experiments are carried out on saturated cylindrical specimens with an initial porosity $\phi_0$ varying between 0.334 and 0.348 (i.e., an average relative density $D_r$ of 90% ) (Yamamuro, 1993). The calibrated model parameters and index properties for Cambria sand are given in Table 2. The model predictions are compared with experimental data for the confining pressure range between 2.11 and 15 MPa in Figure 9. The predicted stress ratio in specimens subjected to different confining pressures evolves toward the critical state stress ratio $M$ under shearing (Figure 9a). Particle breakage is observed in all model simulations (Figure 9c). A good agreement is observed between the stress-strain (Figure 9b), volume change (Figure 9d) and PSD evolution (Figure 9e,f) results from the experiments and those from the model simulations. The model successfully predicts the dilatant response exhibited by the sample subjected to 2.11 MPa confining pressure and suppression of dilatancy by breakage as confinement increases (Figure 9d). Particle crushing, which mostly occurs in the form of progressive comminution of smaller sized fragments, is well captured by the model (Figure 9e and 9f). Overall, the developed model is very successful in predicting the complex response and PSD evolution of two different sands under a wide range of confining stresses.

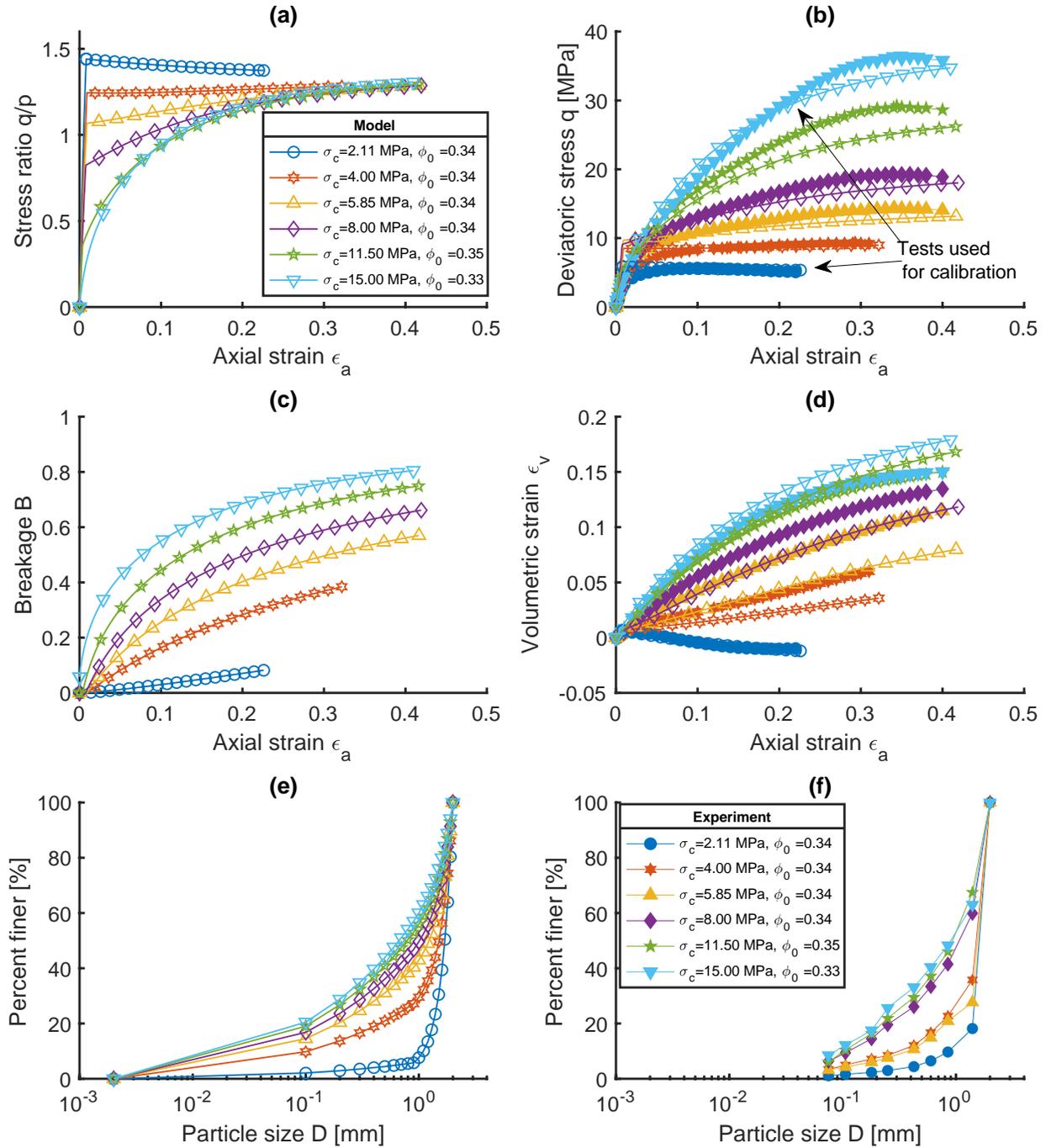

Figure 9. Drained triaxial compression test results (Yamamuro, 1993, Yamamuro and Lade, 1996) and model predictions for Cambria sand under confining pressures between 2.11 MPa and 15 MPa. (a) model prediction of stress ratio versus axial strain response; (b) comparison of the deviatoric stress-axial strain relationship of experiments (curves with solid symbols) with model predictions (curves with open symbols); (c) model prediction of breakage $B$ growth; (d) comparison of the volume change response of experiments (curves with solid symbols) with model predictions (curves with open symbols); (e) model prediction of cumulative PSD; (f) cumulative PSD data measured after experiments.

## 5. Summary and Conclusions

A constitutive model for granular materials composed of breakable particles was developed based on a reformulation of the breakage mechanics theory (Tengattini, *et al.*, 2016). The general formulation of the model, its predictive capabilities, a calibration strategy for model parameters, and the model validation against two sets of experimental data were presented and discussed in detail. The model is shown to capture the main features of the behavior of brittle granular materials at a wide range of confining pressures. In particular, the introduced enrichments allow the proposed formulation to simulate (i) the simultaneous evolution of dilation and breakage that captures the gradual suppression of peak strength and dilation with increasing breakage as confining pressure increases, and (ii) strain softening observed in dense dilatant specimens.

The concurrence of dilation and breakage was modeled by recognizing that particles commence crushing at a specific relative porosity-dependent energy threshold, which can lie within the stress regime causing dilatancy, during shearing. The developed model can qualitatively reproduce the experimentally observed stress-dilatancy-breakage relationship in granular soils at low stresses. A good agreement between the stress-strain, volume change and PSD evolution results of experiments and simulations over a wide range of confining pressures verifies the predictive capabilities of the model. With previously introduced features and new enhancements, the proposed model provides an alternative strategy to predict constitutive behavior and the critical state for brittle granular materials without using the non-linear, pressure- and grading-dependent critical state line as a prerequisite model input.

## 6. Acknowledgements